\documentstyle[floats,multicol,aps,epsf,prb]{revtex}

\newcommand \beq  {\begin{equation}}
\newcommand \eeq  {\end{equation}}
\newcommand \bea {\begin{eqnarray} }
\newcommand \eea {\end{eqnarray}}

\begin{document}
\draft
\twocolumn[\hsize\textwidth\columnwidth\hsize\csname @twocolumnfalse\endcsname
\title{
Sum rule for the optical Hall angle
}
\author{ H. D. Drew$^1$ and P. Coleman$^{2,3}$}
\address{
$^1$Dept of Physics,   
University of Maryland,
College Park, MD 20742. }
\address{
$^2$Serin Laboratory, Rutgers University, P.O. Box 849,
Piscataway, NJ 08855-0849.}
\address{
$^3$ Institute for Theoretical Physics, 
UCSB,
Santa Barbara, CA 93106- 4030.
}
\maketitle
\date{\today}
\maketitle
\begin{abstract}
We consider the optical Hall conductivity of a general
electronic medium and prove that the
optical Hall angle obeys  a new sum rule. This sum rule
governs the response of an electronic fluid to a 
Lorentz electric field and can 
thought of as the transverse counterpart to the 
f-sum rule in  optical conductivity.
The physical meaning of this sum rule is discussed,
giving a number of examples  of its application to a variety of
of electronic media.
\end{abstract}

\vskip 0.2 truein
\pacs{78.20.Ls, 47.25.Gz, 76.50+b, 72.15.Gd}
\newpage
\vskip2pc]

The optical Hall conductivity is a new experimental probe.\cite{orenstein,drew}
Like the optical conductivity, by extending
Hall conductivity measurements into the microwave and far infrared,
it should be possible to extract a host of new information 
about the properties of electronic systems in a magnetic field.\cite{shastry}
Electronic systems where this probe might prove particularly
important, are the cuprate metals,\cite{drew} 
type II superconductors\cite{larkinaronov,vortex_state} and
the integer and fractional quantum Hall systems.\cite{hallref}

One of the most useful tools in the
experimental analysis of the optical conductivity, is the f-sum 
rule\cite{fsum}
\begin{eqnarray}
2\int_0^{\infty}
{d\omega \over \pi} 
\sigma'_{xx}(\omega) = \epsilon_o\omega_p^2
\end{eqnarray}
where 
$\sigma'_{xx}(\omega)$ is the real part of the conductivity
and $\omega_p$ is the plasma frequency.
The distribution of optical spectral weight in an electronic medium
gives us important information about its  underlying physics.
In this paper, we introduce a corresponding sum rule for the 
analysis of the optical Hall angle.
The optical Hall angle, defined as
\begin{eqnarray}
{\rm tan} \theta_H(\omega) =
 {\sigma_{xy}(\omega)/\sigma_{xx}(\omega)},
\end{eqnarray}
where $\sigma_{xx}$ and $\sigma_{xy}$ are the optical and
Hall conductivities respectively, can be measured directly
in optical transmission experiments.\cite{orenstein,drew}
We shall show that this 
response function   
obeys the sum
rule 
\begin{eqnarray}
2\int_0^{\infty}
{d\omega \over \pi} 
{\rm tan} \theta_H'(\omega) = \omega_c, \label{sum}
\end{eqnarray}
where $\omega_c = e B /m$ is the cyclotron frequency of free
electrons. Whereas the f-sum rule governs
the longitudinal response to an applied electric
field, this sum rule governs the transverse 
response to a  Lorentz force.
The transverse sum rule (\ref{sum}) will
be refered to as the ``t-sum" rule. 
The behaviour of the  transverse spectral weight is, in general,
independent of the longitudinal spectral weight: we
shall see how this enables us to make a new, qualitative
distinction between  normal metals,
superconductors, cuprate metals and quantum Hall systems.

The optical Hall angle extends the concept of a Hall angle
to include retardation. 
In linear response theory, an input current $j_y$ leads to 
the retarded  Hall current in the
$x$ direction, $j_x$(Fig. 1. ) as follows
\begin{eqnarray}
j_{x}(t)=\int^t_{-\infty} 
dt'
{\rm tan} \Theta_H(t-t')j_y(t'),\label{retard}
\end{eqnarray}
where ${\rm tan} \Theta_H(t)\equiv{\rm tan} \Theta_H^{(R)}(t)$ is the retarded
 Hall response function.
\begin{figure}[tb]
\epsfxsize=3.5in 
\centerline{\epsfbox{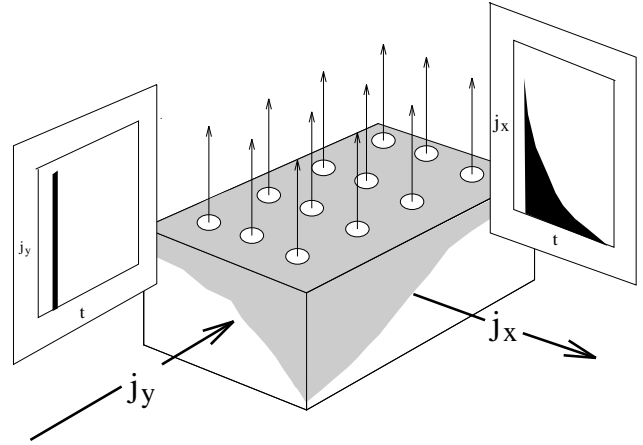}}
\vskip 0.3truein
\protect\caption{
Illustrating the Hall response $j_x(t)= j_o{\rm tan} \Theta_H(t)$ 
to an input current pulse $j_y(t) = j_o\delta(t)$.
In this thought experiment, 
the transverse electric field $E_x$ is
constrained to be zero. }
\label{Fig1}
\end{figure}

Our derivation leans heavily on the analytic properties
of ${\rm tan} \theta_H(\omega)$, which we now discuss.
Since ${\rm tan} \Theta_H(t)$ is a causal
response function, it vanishes for negative times. General
considerations tell us that  its Fourier
transform ${\rm tan} \theta_H(\omega)$, is analytic
in the upper-half complex plane and satisfies the
Kramer's Kr\"onig relations.\cite{landau}
\begin{eqnarray}
\tan \theta_H(\omega+i\delta)
=
\int {dx \over \pi i}
{1 \over (x- \omega - i \delta)}
\times \left\{
\begin{array}{rl}
i \rm tan \theta_H''(x)\cr
 \rm tan \theta_H'(x)
\end{array}
\right.
\label{kronig}\end{eqnarray}
The extension of these two expressions
into the lower-half plane
describe two different Riemann sheets of the function, 
and we must select the physical
sheet that
correctly describes retarded {\sl and}
advanced response functions as follows:
\begin{eqnarray}
{\rm tan} \Theta_H^{(R,A)}(t)= \int {d \omega\over 2 \pi}
 {\rm tan} \theta_H(\omega\pm i\delta) e^{-i \omega t}.\label{ftrans}
\end{eqnarray}
Since 
$\sigma^{(R)}_{xx}(t)=- \sigma^{(A)}_{xx}(-t)$ and 
$\sigma^{(R)}_{xy}(t)=\sigma^{(A)}_{xy}(-t)$
are respectively odd and even under time-reversal, 
${\rm tan} \theta_H= \sigma_{xy}\otimes \sigma_{xx}^{-1}$
is odd under time-reversal. Taking the Fourier
transform
(\ref{ftrans}),
it follows that 
${\rm tan} \theta_H(\omega)^*=-{\rm tan} \theta_H(\omega^*)$,
i.e, the physical Riemann sheet has 
a cut along the real axis where only the
real-part of ${\rm tan} \theta_H(\omega)$ changes sign.  
The extension of the second Kramer's Kr\" onig
relation (~\ref{kronig}) into the complex plane
\begin{eqnarray}
{\rm tan} \theta_H(z) = \int_{-\infty}^{\infty} {d \omega \over  \pi i}
{1 \over \omega-z}
{\rm tan} \theta_H'(\omega).\label{spectral}
\end{eqnarray}
provides the Riemann sheet 
that is consistent with this branch-cut structure.
This spectral representation for the Hall angle holds
on general grounds, {\sl even though}
we do not have an explicit  Kubo formula for the Hall angle.

We now present   a physical derivation of the t-sum rule.
Suppose a small  current pulse 
$j_y(t) = j_o \delta (t)$ is applied to the system.
During the pulse, the current
will precess
at a rate $\omega_c$ about the magnetic field. The important
point here, is that provided the 
pulse is brief enough, interactions and band-structure
effects are negligible, precession is determined by the 
free-field Hamiltonian $H_o$, so that  
\begin{eqnarray}
 \dot { \vec j} = i [H_o, \vec j]=-\omega_c  \hat z \times \vec j,
\end{eqnarray}
where $\hat z$ is the direction of the magnetic field.
The transverse Hall current that builds up during the
pulse is 
\begin{eqnarray}
j_x(0^+) = \omega_c\int_{0^-}^{0^+} dt j_y(t)  = \omega_c j_o.
\end{eqnarray}
By (\ref{retard}), 
$
j_x(0^+)= {\rm tan} \theta_H(0^+)j_o $ so that 
\begin{eqnarray}
{\rm tan} \theta_H(0^+)  = \omega_c =
\int _{-\infty}^{\infty} {d \omega \over 2 \pi} 
{\rm tan} \theta_H(\omega+i\delta)e^{-i\omega 0^+}
\end{eqnarray}
By extending this integral along an infinite semi-circular contour 
in the lower-half complex plane and deforming the contour
onto the cut along the real axis,  where 
${\rm tan} \theta_H(\omega)$ 
has a discontinuity, the t-sum rule (\ref{sum}) follows.

A more general derivation follows from  the Kubo formula.
Let us take the magnetic field to lie along the z-direction,
assuming the conductivity to be isotropic about this axis.
The conductivity tensor in the basal plane 
can be written in the  form
$\sigma_{ab}(\omega)
= Q_{ab}(\omega)/(-i\omega)$, ($a,\ b = x,y$)
where in the time domain, 
\begin{eqnarray}
Q_{ab}(t) =
 \epsilon_o \omega_p^2\delta_{ab}\delta(t)
&-& i  \langle [\hat j_a(t),\hat j_b(0)]\rangle \theta(t).
\end{eqnarray}
The delta-function part is 
the instantaneous diamagnetic response, which is  
diagonal by inversion symmetry. 
The dominant short-time behavior of $Q_{ab}(t)$ is given by
\begin{eqnarray}
\left.
\begin{array}{rcl}
Q_{xx}(t) &\sim&  \epsilon_o \omega_p^2\delta(t), \cr
Q_{xy}(t) &\sim& - i  \langle [\hat j_x,\hat j_y]\rangle 
\theta(t),
\end{array}\right\}\qquad  ( t\rightarrow 0).
\end{eqnarray}
The Fourier transform of these expressions determine the
asymptotic high frequency behavior of the conductivity tensor, so that
\begin{eqnarray}
\begin{array}{rcl}
\sigma_{xx}(z) &=& 
\displaystyle {\epsilon_o \omega_p^2 / (-i z) }, 
\cr
\sigma_{xy}(z) &=& \displaystyle {i \langle [\hat j_x, \hat j_y]
\rangle / z^2 },
\end{array}
\qquad\qquad(|z| \rightarrow \infty).
\end{eqnarray}
The asymptotic form of the Hall conductivity 
was previously obtained by Shraiman and Shastry,
using a spectral decomposition.\cite{shastry}
Taking the ratio, we obtain ${\rm tan} \theta_H(z) = 
\omega_H/(-iz)$, ($|z| \rightarrow \infty$), where 
\begin{eqnarray}
\omega_H= \left( {-i \langle [\hat j_x, \hat j_y]
\rangle \over \epsilon_o \omega_p^2 }\right).
\end{eqnarray}
The Hall frequency, $\omega_H$ is the total residue of ${\rm tan} \theta_H(z)$
along the
real axis. Multiplying the spectral representation (~\ref{spectral})
by $-iz$ and taking $\vert z \vert\rightarrow \infty$, 
we obtain
\begin{eqnarray}
2\int_0^{\infty}
{d\omega \over \pi} 
{\rm tan}\theta_H'(\omega) = 
\omega_H.
\end{eqnarray}

For an electron gas  with quadratic dispersion
$
 -i  [\hat j_x, \hat j_y]  = \omega_c (\omega_p^2 \epsilon_0),
$
so $\omega_H = \omega_c$.
We will give a derivation of this
result momentarily. 
Since all electronic systems are ultimately derived from
a system with a quadratic dispersion, this is an exact result, but 
recovery of the full spectral weight   requires
an integration over all inter-band transitions.  
The usefulness of the sum
rule derives from the fact that when the bands
are well separated,  an effective sum rule  applies  to the lowest
band.
Consider  a single  band described by the 
Hamiltonian
\begin{eqnarray}
H[A] = \sum_{\vec p } \epsilon_{\vec p- e \vec A}\hat n_{\vec p},
\end{eqnarray}
where $\hat n_{\vec p}= \psi^{\dagger}_{\vec p}\psi_{\vec p}$
is the number operator.
The plasma frequency of the band is given by 
\begin{eqnarray}
\epsilon_o \omega_p^2 = \langle \nabla^2_{\vec A_x} H[A] \rangle
= (e^2/2) \sum_{\vec p } {\rm Tr}
 (
\underline{m} _{ \rm \vec p}^{-1})
n_{\rm \vec p}\label{one}
\end{eqnarray}
where $[\underline{m}_{\rm \vec p}^{-1}]_{ab} = \nabla^2_{ab}
\epsilon_{\rm \vec p }$ is the two-dimensional effective mass
tensor and $n_{\vec p} = \langle \hat n_{\vec p}\rangle$. 
Provided that the magnetic flux per unit cell is far less than a flux
quantum $h/e$, we may use a weak field approximation to
$[\hat j_x, \hat j_y]$, obtained by linearizing the velocity operator
as follows
\def\pbar{{  p\kern-1.1ex\raise-0.2ex\hbox{/}}}
\def\vpbar{{  \vec p\kern-1.1ex\raise-0.2ex\hbox{/}}}
\begin{eqnarray}
\vec v_{\vec p - e \vec A}
= \vec v_{\vec p_o} 
+\underline{m}^{-1}_{\vec p}
({{  \vec p\kern-1.1ex\raise-0.2ex\hbox{/}}}-\vec p_o),
\end{eqnarray}
where ${{  \vec p\kern-1.1ex\raise-0.2ex\hbox{/}}} = \vec p - e \vec A$.
Since 
$
[{  p\kern-1.1ex\raise-0.2ex\hbox{/}}_x, {  p\kern-1.1ex\raise-0.2ex\hbox{/}}
_y] = -i eB_z
$, 
it follows that
$
[\vec v_x, \vec v_y] = 
{(eB) {\rm det}(\underline {m}_{\rm \vec p}^{-1})} 
$.
Writing $\hat j = \sum \psi^{\dagger}_{\rm \vec p}\vec v_{{{  \vec p\kern-1.1ex\raise-0.2ex\hbox{/}}}}
 \psi_{\rm \vec p}$, we obtain the operator identity
\begin{eqnarray}
[\hat j_x,\hat j_y] = 
\sum_{\vec p}
\left({e^3  B{\rm det}(\underline {m}_{\rm \vec p}^{-1})} \right)
\hat n_{\vec p}+ O(B^3)\label{two}.
\end{eqnarray}
Combining (\ref{one}) and (\ref{two}), it follows that 
\begin{eqnarray}
\omega_H =  {e B} {\sum_{\vec p}   {\rm det}
[\underline{m}^{-1}_{\vec p}] n_{\vec p}
\over \frac{1}{2}\sum_{\vec p} 
 {\rm Tr}[\underline{m}^{-1}_{\vec p}] n_{\vec p}}.\label{cycl}
\end{eqnarray}
For a parabolic band, 
(~\ref{cycl}) reverts to
the   free cyclotron frequency $\omega_c= eB/m$. Since  
this expression is dominated by contributions far
from the Fermi surface, the Hall frequency
will be only weakly temperature dependent.
Suppose the system
possesses a Fermi surface where
$n_{\rm p} =1$ far inside  and $n_{\rm p} =0$
far outside.  To extract the dominant contribution to $\omega_H$, 
we  we may set $n_{\rm \vec p}=1$ and 
restricting  the integrals within each Fermi surface sheet.
Both integrands are total derivatives,
\begin{eqnarray}
{\rm det}[\underline{m}^{-1}_{\vec p}] &=& 
-\frac{1}{2} \nabla\times\bigl[
(\nabla\vec u) \times \vec u \bigr],
\cr
{\rm Tr}[\underline{m}^{-1}_{\vec p}] &=&  \nabla \cdot \vec u,
\end{eqnarray}
where $\vec a \times \vec b = \epsilon_{\alpha \beta}a_{\alpha}b_{\beta}$
denotes the two-dimensional cross-product, $\vec u$ denotes the
component of the group velocity in the x-y plane and 
$(\nabla\vec u)_{ab} = \underline{
\rm m}^{-1}_{ab}$. This
enables us to re-write the t-sum rule as a ratio of 
Fermi surface integrals.
\begin{eqnarray}
\omega_H = e B  {\int dp_z \int_{\rm FS}\vec u\times d \vec u\over
\int dp_z \int_{\rm FS} d\vec S \cdot \vec u } +O (\delta \epsilon^2/
\epsilon_F^2).
\end{eqnarray}
Here ``FS'' denotes a line integral around all sheets
of the Fermi surface
at constant $p_z$, $d\vec S$ denotes the surface increment
that lies perpendicular to this line,
$d \vec u= dk.\nabla \vec u$ is the change in $\vec u$ along
the line and $\delta \epsilon/\epsilon_F$ is
the ratio of the smearing of the Fermi surface to the
Fermi energy.
The numerator is readily identified as twice
the area swept out by the Fermi velocity
$\vec u$ around the Fermi surface.
A variant of this  expression has been obtained by Ong\cite{ong}
using Boltzmann transport theory.
Provided the Fermi surface
is reasonably well-defined, we expect this expression to be 
robust.

We now illustrate the qualitative implications of the Hall sum rule
using a few physical examples. In the case of a simple metal,
the transverse optical conductivity in a magnetic field is given by
\begin{eqnarray}
\sigma_{\pm}=\sigma_{xx}\pm i \sigma_{xy} =     {ne^2 \over m}{1 \over \Gamma_{tr}
-i(\omega\mp \omega_c) }\label{drude}
\end{eqnarray}
so that the optical conductivity is peaked at the cyclotron frequency.
Remarkably, these poles do not enter the Hall response, 
\begin{eqnarray}
{\rm tan} \theta_H(\omega) = { \omega_c \over \Gamma_{tr} - i \omega },
\end{eqnarray}
which has the same form as the zero field optical conductivity (Fig. 2a).
The Hall spectral weight is peaked at zero frequency and is independent
of carrier density.   In this case, 
the Hall constant 
$R_H(\omega) = {\rm tan} \theta_H(\omega)/\sigma_{xx}(\omega)$ is frequency
independent.  

An intriguing exception to this behavior
is found in  the cuprate metals.
Unlike conventional metals, 
transport measurements indicate that $\Gamma_{tr}
\sim 2T$, but 
D.C. Hall measurements show that $[{\rm cot} \theta_H] \sim  T^2 $
is a quadratic function of temperature.
Since the
D.C. Hall angle scales as $1/T^2$, the sum rule tells
us that the Hall
relaxation rate $\Gamma_H$ is a quadratic function of tempterature
$\Gamma_H = T^2/W$ and that furthermore
$\sigma_{xy}\sim 1/(\Gamma_{tr}\Gamma_H)$. 
This multiplicative combination of relaxation rates
is unprecedented and 
does not fit into a conventional picture of normal metals.
Based on this result, Anderson has conjectured
that the Hall currents are subject to
an autonomous decay process that depends quadratically on temperature
$\Gamma_H \propto T^2$.\cite{phil}
This controversial interpretation 
follows  naturally from a Hall sum rule, in a fashion that is
independent of the microscopic physics.  
A definitive measurement of the quadratic temperature
dependence of the Hall decay rate 
would constitute  a striking confirmation of
the power of the Hall sum rule. (Fig. 2(b))

\begin{figure}[tb]
\epsfysize=5.in 
\centerline{\epsfbox{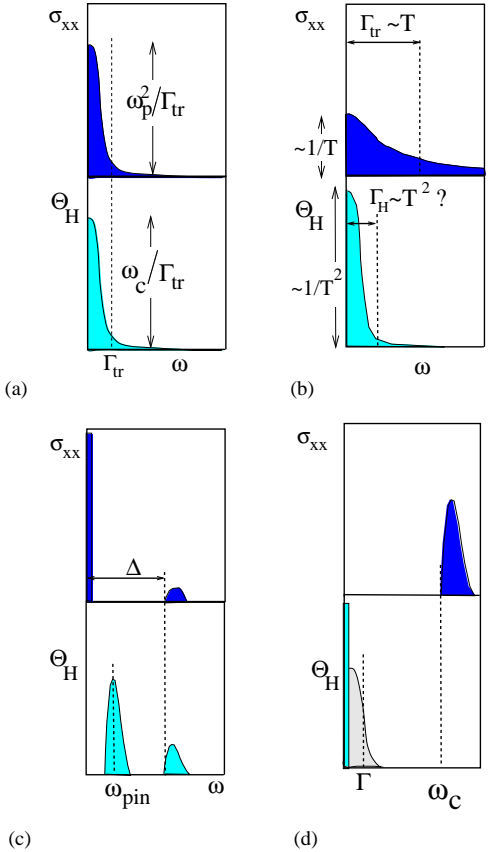}}
\protect\caption{
Contrasting the optical conductivity and the optical Hall angle
in (a) a simple metal (b) a cuprate metal,  (c) a superconductor and 
(d) a quantum Hall fluid. The superconductor displays condensation
in the dynamical conductivity; the quantum Hall fluid displays
displays a condensation in the optical Hall angle at a Hall plateau
which periodically broadens into a Drude peak between plateaus. 
}
\label{Fig2}
\end{figure}

As a second  application, 
consider a type
II superconductor with a pinned vortex lattice.
In a superconductor, the optical conductivity
condenses into a zero-energy delta function peak.
However, the same condensation does not take place in the
Hall angle, because unlike conventional
currents, super-currents can not precess in a magnetic field:
the deflection of a supercurrent
requires a sideways movement of the flux lattice. 
Hsu\cite{tedhsu}
has computed the Hall response of a pinned flux lattice, and shown that
it is shifted to the flux lattice pinning frequency (Fig. 2(c)), as 
observed in recent experiments on YBCO\cite{lihn}.
Pehaps the most important property of the sum rule in this respect,
is that it does not depend on temperature or the thermodynamic state
of the system.  Despite the
radically different physics of the vortex lattice and the normal
state, the Hall sum is identical.

As a final example, we consider the Hall response of a two-dimensional electron
gas in a high magnetic field. 
When the quantized, or fractionally quantized Hall ground-state develops,
the conductivity (~\ref{drude}) is qualitatively modified, leading to the
quantum Hall effect in the D.C. response : ${\sigma_{xx}}=0$, 
$\sigma_{xy}= \nu e^2/h$,
where $\nu=p/q$ is a
rational number with an odd denominator.
However, the oscillator strength sum rule is still dominated
by the poles at $\omega_c$. What happens to the Hall angle?
Like an  insulating dielectric, the
optical conductivity at a Hall plateau
vanishes linearly with frequency
$\sigma_{xx}(\omega) = \alpha (-i\omega) $ at $T=0$.\cite{kallin}
This implies that the a.c. Hall angle has the form
\begin{eqnarray}
{\rm tan} \theta_H(\omega) = {1 \over -i \omega}
\left({\nu e^2\over h\alpha}\right)
\end{eqnarray}
i.e. the Hall angle response has condensed into a delta function.
Assuming all the Drude weight condenses, then ${\rm tan} \theta_H(\omega) = 
{\omega_c / (-i \omega)}$ and $\alpha$ takes its minimum value
$\alpha = \nu e^2/h \omega_c$.
From this qualitative reasoning, we clearly see how
 a quantum Hall system at $T=0$ 
is the transverse  counterpart to a superconductor.
In a superconductor,
the longitudinal current accelerates in response to an electric field, 
but there is no  Hall response.  In a quantum Hall system, the 
transverse Hall current exhibits a superfluid
response to a Lorentz electric field, but
there is no longitudinal  response. 
Between Hall steps, we expect the delta function to broaden into
a Drude form of width $\sim \Gamma_{tr}$.  At $T=0$ this
occurs abruptly as a function of electron density or magnetic
field, corresponding to a quantum phase transition.  
Therefore, we again see the analogy between the Hall angle
in this system and and the change in the conductivity at
a superconducting phase transition.
At finite temperatures, $\sigma'_{xx}\sim {e^2 \over\hbar}
e^{-E_g/{k_B T}}$, where $E_g$ is the gap in the density of states, so
we expect the delta-function to broaden into a Lorentzian  of width
$\delta \omega \sim \nu \omega_c e^{-E_g/{k_B T}}$.
The Hall angle sum rule should prove very useful in studies of
the conductivity of quantum Hall systems,
since it gives a spectral sum rule that may saturate at
Frequencies $\omega \sim \Gamma_{tr}<< \omega_c$, which is
the physically
interesting range of frequencies.

In conclusion, we have considered the
optical Hall angle as a dynamical response function,
and showed that it obeys a  sum rule that governs
the evolution of transverse Hall currents.
From our 
 discussion of its qualitative application to 
metals, cuprate metals, BCS superconductors and quantized Hall systems,
we see that the optical Hall angle may be thought of as the
transverse analogue to the optical conductivity.
Like the f-sum rule of the optical conductivity, the corresponding
t-sum
rule is independent of
detailed  microscopic physics, making it of 
great utility in the qualitative
analysis of magneto-optic response of electronic systems. 

We should like to thank R. Shankar, A. J. Schofield and
S. C. Zhang  for discussions related to this work.   
This work was supported in part by the National Science Foundation
under Grants numbers DMR-93-12138,   DMR-92-23217 and PHYS94-07194 
during the course of a stay at the ITP, Santa Barbara.

\end{document}